\documentclass[twocolumn,showpacs,prl,aps]{revtex4}

\usepackage{dcolumn}

\draft

\begin{document}

\newcommand{\Tr}{\texttt{Tr}}

\title{The Schr\"{o}dinger Equation for Open Systems}

\author{Yuriy E. Kuzovlev}
\email{kuzovlev@kinetic.ac.donetsk.ua}
\affiliation{A.A.Galkin Physics and Technology Institute
of NASU, 83114 Donetsk, Ukraine}


\begin{abstract}
An universal exact description of kinetics of open quantum systems in
terms of random wave functions and stochastic Schr\"{o}dinger
equation is suggested. It is shown that evolution of random quantum
states of an open system is unitary on average, and this implies
validity of the optical theorem for any inelastic scattering.
\end{abstract}


\maketitle

{\bf 1}. As it is known, one not always can attribute to a quantum
system some definite wave function. Instead a density matrix is
necessary \cite{ll}. Anyway if the matter concerns an open system
$\,\mathcal{S}\,$ which is a part of more complicted closed system
$\,\mathcal{S}$+$\mathcal{W}\,$ (where $\,\mathcal{W}\,$ is, for
instance, ``thermal bath'' or ``the rest of world''). In such a case
even density matrix of $\,\mathcal{S}\,$ is not quite definite
because it is impossible to write out an exact general equation for
it. In practice one is supposed to manage with ``kinetic equations''
deduced either from the von Neumann equation for the whole system
$\,\mathcal{S}$+$\mathcal{W}\,$ or from speculative reasonings
\cite{lp,bp,uw,parth}. Sometimes irreversible terms there are
supplemented by noise terms to be averaged later. Somehow or other,
usually one at first constructs approximate equations and then
constructs a method of their approximate solving.

Meanwhile one may avoid such duplication of approximations if recruit
``stochastic representation of quantum interactions'' \cite{i2} (see
also \cite{a1,a2,i4}). The latter produces a closed stochastic
equation for $\,\mathcal{S}\,$ which is extremely simple, intuitively
obvious and at the same time formally exact:\, if random sources in
it reflect, by definite universal rules, internal dynamics of
$\,\mathcal{W}\,$, then such sources act exactly as actual dynamic
interaction of $\,\mathcal{S}$ with $\mathcal{W}\,$. Importantly,
these random sources introduce at once noise and irreversibility as a
statistical effect of the noise. Therefore approximation is required
only once, under calculation of statistical averages. Related
mathematical problems are similar to problems of theory of
oscillations and waves in linear media with fluctuating parameters
\cite{kl}.

A purpose of the present article is to emphasize possibility to
reformulate the approach suggested in \cite{i2} (see also
\cite{i4,i5,i6,i7}) in terms of stochastic Schr\"{o}dinger equation
and thus generalize to open systems the concepts of pure quantum
state and wave function. Correspondingly, one can apply to open
systems, in an elementary statistical sense, the concept of unitarity
of evolution which naturally coexists with irreversible evolution of
averaged wave function and density matrix of $\,\mathcal{S}\,$. As
the result, the problem of calculation of its averaged (that is true)
density matrix can be substantially simplified. At that, an important
roles is played by average wave function.

\,\,\,

{\bf 2}. We will start with brief summary of the stochastic
representation, following \cite{i2} (with not great difference of
designations) and \cite{i7}. Let us divide the whole system
Hamiltonian into three parts and write a part $\,H_{int}\,$
responsible for the interaction as a bilinear form:
\begin{equation}
\begin{array}{c}
H\,=\,H_S+H_W+H_{int}\,\,\,,\,\,\,\,\,\, H_{int}\,=\,\sum_j
S_j\,W_j\,\,\,,\label{ham}
\end{array}
\end{equation}
where operators $\,S_j\,$ и $\,W_j\,$ act in different Hilbert spaces
of $\,\mathcal{S}\,$ and $\,\mathcal{W}\,$, respectively. Such
decomposition of $\,H_{int}\,$ frequently arises from very physical
nature of the interaction and in any case is feasible in formal sense
\cite{bp}. Then for any operator $\,\mathcal{O}\,$ introduce
Liouville super-operator $\,\mathcal{L}(\mathcal{O})\,$ and Jordan
super-operator $\,\Pi(\mathcal{O})\,$ as follow:
\[
\mathcal{L}(\mathcal{O})A\equiv\frac
i{\hbar}\,(A\mathcal{O}-\mathcal{O}A)\,\,,\,\,\,
\Pi(\mathcal{O})A\equiv\frac 12
\,(A\mathcal{O}+\mathcal{O}A)\,\,,\nonumber
\]
where $\,A\,$ is arbitrary operator. Thus\, $\,\mathcal{L}(H)\,$ is
Liouville super-operator in the von Neumann equation $\,\dot\rho
=\mathcal{L}(H)\rho\,$ for density matrix $\,\rho\,$ of the whole
system $\,\mathcal{S}$+$\mathcal{W}\,$. One can easy verify that
correspondingly to (\ref{ham}) it expands to
\begin{eqnarray}
\mathcal{L}(H)\,= \,\mathcal{L}(H_S)+\mathcal{L}(H_W)+
\,\,\,\,\label{l}\\
+\sum\,[\,
\mathcal{L}(S_j)\,\Pi(W_j)+\Pi(S_j)\,\mathcal{L}(W_j)\,]\nonumber
\end{eqnarray}
Therefore, density matrix\, $\,\rho_S(t)\,=\,\Tr _{\,W}\,\rho
(t)\,$\, of $\,\mathcal{S}$ can be expressed by
\begin{equation}
\begin{array}{c}
\rho_S(t)\,=\,\Tr_{\,W}\,\,\overleftarrow{\exp }\left[\,\int^t_0
\mathcal{L}_S(t^{\,\prime\,})
\,dt^{\,\prime}\right]\rho(0)\,\,\,,\label{rs}
\end{array}
\end{equation}
where\, $\,\overleftarrow{\exp }\,$ is chronologically ordered
exponential,
\begin{eqnarray}
\mathcal{L}_S(t)\,\equiv \mathcal{L}(H_S)+\sum
\,[\,x_j(t)\,\mathcal{L}(S_j)+ y_j(t)\,\Pi(S_j)\,]\,\,\,,\label{xys}
\end{eqnarray}
and\, $\,x_j(t)\,$ and $\,y_j(t)\,$ are time-varying super-operators,
\[
\begin{array}{c}
\,\,x_j(t)\,=\, e^{-\,\mathcal{L}(H_W)\,t}\,\Pi(W_j)\,
e^{\,\mathcal{L}(H_W)\,t}\,\,\,,\\
\,\,\,y_j(t)\,=\, e^{-\,\mathcal{L}(H_W)\,t}\,\mathcal{L}(W_j)\,
e^{\,\mathcal{L}(H_W)\,t}\,\,\,,
\end{array}
\]
whose action is completely localized in the $\mathcal{W}\,$'s Hilbert
space. The latter circumstance, if considered from the viewpoint of
$\mathcal{S}\,$, allows us to treat $\,x_j(t)\,$ and $\,y_j(t)\,$ in
(\ref{rs}) and (\ref{xys}) merely like arbitrary commutative (scalar)
time-dependent variables. In other words, like ``noises'' or random
processes. At that, the trace operation in (\ref{rs}) plays the role
of statistical averaging in respect to the random processes
$\,x_j(t)\,$ and $\,y_j(t)\,$.

In order to define effective statistical properties of $\,x_j(t)\,$
and $\,y_j(t)\,$ in a simple unambiguous way, it is convenient to
assume that before some initial time moment $\,t_0\,$, e.g.
$\,t_0=0\,$, the subsystems $\,\mathcal{S}\,$ and $\,\mathcal{W}\,$
were non-interacting and statistically independent one on another:\,
$\,\rho(t_0)=\rho_{S}^{(in)}\rho_W^{(in)}\,$. Then one can write
\begin{equation}
\begin{array}{c}
\rho_S(t)\,=\,\langle\,R(t)\,\rangle\,\,\,,\,\,\,\,\,
\,\dot{R}(t)\,=\,\mathcal{L}_S(t)R(t)\,\,\,,\label{sle}
\end{array}
\end{equation}
where $\,R(t)\,$ is ``random density matrix'' of $\,\mathcal{S}\,$,
which obeys ``stochastic Liouville equation'' with initial condition
$\,R(t_0)=\rho_{S}^{(in)}\,$,\, and angle brackets symbolize the
averaging:\,
\[
\begin{array}{c}
\langle\,...\,\rangle\, =\,\Tr_{\,W}\,...\,\,\rho_W^{(in)}\,\,
\end{array}
\]
On right-hand side here $\,x_j(t)\,$ and $\,y_j(t)\,$ are treated as
the above super-operators while on left side as effective random
processes. This is just the statistical definition of these
processes. Evidently, it means that characteristic functional of
random processes $\,x_j(t)\,$ and $\,y_j(t)\,$ can be presented by
expressions
\begin{eqnarray}
\langle \,\,\exp \int \sum\,[\,u_j(t)\,x_j(t)+f_j(t)\,
y_j(t)\,]\,\,dt\,\,\rangle\,\,=\label{cf0}\\
=\,\Tr_{\,W}\,\,\overleftarrow{\exp } \left[\,\int
\mathcal{L}_W(t)\,dt\,\right]\rho_W^{(in)} \,\,\,, \,\,\,\,\,
\,\,\,\,\,\nonumber
\end{eqnarray}
where $\,u_j(t)\,$ and $\,f_j(t)\,$ are arbitrary probe functions,
\[
\begin{array}{c}
\mathcal{L}_W(t)\,\equiv\,\mathcal{L}(H_W) +\sum\,[\,u_j(t)\,
\Pi(W_j)+f_j(t)\,\mathcal{L}(W_j)\,]\,\,,
\end{array}
\]
the integration begins at $\,t=t_0\,$ and by request one may move
$\,t_0\,$ away to $\,-\,\infty\,$.

According to (\ref{xys}) и (\ref{cf0}), the processes $\,x(t)\,$
represent perturbation of $\,\mathcal{S}\,$ by $\,\mathcal{W}\,$ as
directly prescribed by their Hamiltonian interaction. Analogously,
$\,y(t)\,$ represent back action of $\,\mathcal{S}\,$ onto
$\,\mathcal{W}\,$. In view of these observations, we can expect that
$\,x(t)\,$ are quite usual random processes while, in opposite,
$\,y(t)\,$ are rather unusual ones. Indeed, at $\,u(t)=0\,$ the
super-operator $\,\mathcal{L}_W(t)\,$ reduces to a Liouville
super-operator, and then from (\ref{cf0}) it follows that\,
$\,\langle\, y(t_1)\,...\,\,y(t_n)\,\rangle =0\,$\,. By the same
reason, as we can see from (\ref{cf0}),\, $\,\langle
\,y(t)\,\,x(\tau_1)\,...\,\,x(\tau_m)\,\rangle \,=\,0\,$\, if\,
$\,t>\max_{\,k}(\tau_k)\,$. But generally\, $\,\langle
\,x(\tau_1)\,...\,\,x(\tau_m)\,\,y(t_1)\,...\,\,y(t_n)\,\rangle
\,\neq \,0\,$\, if\, $\,\max_{\,k}(\tau_k)> \max_{\,k}(t_k)\,$.\,
Hence,\, $\,y(t)\,$ show their worth in conjunction with $\,x(t)\,$
only. At that, cross correlations of $\,y(t)\,$ with $\,x(t)\,$
represent response of $\,\mathcal{W}\,$ to its perturbation by
$\,\mathcal{S}\,$ and eventually introduce to evolution of
$\,\rho_S(t)\,$ dissipation and irreversibility. The condition
$\,\max_{\,k}(\tau_k)> \max_{\,k}(t_k)\,$ comes from the
chronological ordering in (\ref{rs}) and expresses the causality
principle. Details and generalizations of such ``stochastic
representation of dynamic interactions'' \cite{i4} can be found in
\cite{i2,a1,a2,i4,i5,i6,i7,i8}.

\,\,\,

{\bf 3}. Now let us go to original theme of this paper.

If initially (before interaction between subsystem under interest
$\,\mathcal{S}\,$ and ``the world'' $\,\mathcal{W}\,$ was switched
on) $\,\mathcal{S}\,$ was in a pure state,
$\,\rho_{S}^{(in)}=|\Psi^{(in)}\rangle\langle \Psi^{(in)}|\,$, then
solution to equation (\ref{sle}) is merely\,
$\,R(t)=|\Psi(t)\rangle\langle \Psi(t)|\,$,\, where $\,\Psi(t)\,$
satisfies ``stochastic Schr\"{o}dinger equation''\,:
\begin{equation}
\frac {d \Psi(t)}{dt}\, =-\frac i{\hbar}\,[\,H_S+\sum\,w_j(t)\,
S_j\,]\,\Psi(t)\,\,\label{sse}
\end{equation}
with initial condition\, $\,\Psi(t_0)=\Psi^{(in)}\,$\, and
\[
\begin{array}{c}
w_j(t)\,\equiv\,x_j(t)\,+\,i\hbar\, y_j(t)/2\,
\end{array}
\]
being interpreted as complex random processes. In such sense,
$\,\mathcal{S}\,$ at all times remains in a ``random pure state''.
Consequently, at arbitrary initial density matrix we can make its
diagonalization,\, $\,\rho_{S}^{(in)}=\sum_{\alpha
}\,|\Psi^{(in)}_{\alpha }\rangle\,P_{\alpha }\,\langle
\Psi^{(in)}_{\alpha }|\,$,\, and then write
\begin{equation}
\begin{array}{c}
R(t)=\,\sum_{\alpha} |\Psi_{\alpha}(t)\rangle\,
P_{\alpha}\,\langle\Psi_{\alpha }(t)|\,\,\,,\label{expan}
\end{array}
\end{equation}
where each of $\,\Psi_{\alpha }(t)\,$ satisfies (\ref{sse}) under
initial condition\, $\,\Psi_{\alpha }(t_0)=\Psi_{\alpha }^{(in)}\,$,
while the weights $\,P_{\alpha }\,$ stay constant.

Because of complexity of $\,w_j(t)$ an evolution described by
equation (\ref{sse}) is not unitary:
\[
\frac {d}{dt}\,\langle\Psi_{\alpha}(t)|\Psi_{\beta}(t)\rangle
\,=\,\sum\, y_j(t)\,\langle \Psi_{\alpha}(t)
|S_j|\Psi_{\beta}(t)\rangle\,\neq\, 0\,
\]
Nevertheless, it is unitary on average:
\[
\frac {d}{dt} \,\left\langle\,
\langle\Psi_{\alpha}(t)|\Psi_{\beta}(t)\rangle \,\right\rangle
=\langle\sum\, y_j(t)\langle \Psi_{\alpha}(t)|S_j|\Psi_{\beta}(t)
\rangle\,\rangle = 0
\]
This statement directly follows from the above mentioned statistical
properties of $\,y(t)\,$ (that is factually from the causality
principle). As the consequence,
\begin{equation}
\begin{array}{c}
\left\langle\,\, \langle \Psi_{\alpha}(t)| \Psi_{\beta}(t)\rangle\,
\,\right\rangle = \langle \Psi_{\alpha}^{(in)}|
\Psi_{\beta}^{(in)}\rangle\,=\, \delta_{\alpha\beta}\,\,\,,\label{au}
\end{array}
\end{equation}
and one can say that the ``random pure states'' all the time are
mutually orthogonal on average.

Let us suppose that the interaction by its nature is localized at
time, that is has character of scattering process. In such a case,
random time-varying state of $\,\mathcal{S}\,$ can be written as sum
\[
\begin{array}{c}
|\Psi(t)\rangle\, =\,
|\Psi^{0}(t)\rangle\,+\,|\Psi^s(t)\rangle\,\,\,,
\end{array}
\]
where $\,\Psi^{0}(t)\,$ represents free evolution of
$\,\mathcal{S}\,$ while $\,\Psi^s(t)\,$  is contribution of the
scattering, so that\, $\,\Psi^s(t_0)=0\,$. Then, due to the unitarity
on average, as well as unitarity of the free evolution, the equality
\[
\begin{array}{c}
\langle\,\langle\Psi(t)|\Psi(t)\rangle \,\rangle\,
=\,\langle\Psi^0(t)|\Psi^0(t)\rangle
\end{array}
\]
takes place. It implies that
\begin{equation}
\begin{array}{c}
2\,\texttt{Re}\, \langle\,\Psi^{0}|\langle
\,\Psi^s\,\rangle\rangle\,+\, \langle
\,\langle\,\Psi^s|\Psi^s\rangle\,\rangle \,=\,0\,\,\label{sau}
\end{array}
\end{equation}
This is generalized ``optical theorem'' which is valid for arbitrary
non-elastic scattering.

This result demonstrates that the average wave function $\,\langle
\,\Psi(t)\,\rangle\,$ contains несет important information about an
open system. It should be underlined that $\,\langle
\,\Psi(t)\,\rangle\,$ is always unambiguously defined by equation
(\ref{sse}) as soon as random processes $\,w(t)\,$ are defined by
(\ref{cf0}). Therefore, the average wave function is not less
legitimate than the density matrix $\,\rho_{S}(t)\,$.

Notice also that if $\,w(t)\,$ in (\ref{sse}) are stationary random
processes then average wave functions  $\,\langle \Psi^s(t)\rangle\,$
in (\ref{sau}) and $\,\langle \Psi(t)\,\rangle=\Psi^0(t)+\langle
\Psi^s(t)\rangle\,$ include those energies and frequencies only what
already are present in $\,\Psi^{0}(t)\,$. Therefore non-elastic
component of the scattering wholly comes from fluctuations\,
$\,\Psi(t)-\langle \Psi(t)\rangle\,$, although the latter contribute
to elastic component too. In view of this circumstance, relations
(\ref{au}) or (\ref{sau}) imply inequality
\[
\begin{array}{c}
\mathcal{P}^{\,inel}(t)\,\leq\,1-\langle
\,\langle\Psi(t)\rangle\,|\,\langle\Psi(t)\rangle\,\rangle\,\,\,,
\end{array}
\]
with $\,\mathcal{P}^{\,inel}(t)\,$ being total probability of
non-elastic scattering. In case of a scattering center in
one-dimensional conducting channel one can obtain additional
inequality $\,\mathcal{P}^{\,inel}\leq 1/2 \,$\, \cite{i5,i7}.

{\bf 4}. The stationarity condition realizes, for instance, when
$\,\rho_{W}^{(in)}\propto\exp{(-\,H_W/T)}\,$, that is subsystem
$\,\mathcal{W}\,$ initially was thermodynamically equilibrium (all
the more when $\,\mathcal{W}\,$ is a thermostat). Then
\cite{i2,i4,a1,i5,i7,i8}
\[
\begin{array}{c}
\langle w^{*}_j(\tau )\,, w _m(0)\rangle \,= \,K_{jm}(\tau )
\,\,\,,\,\, \\ \langle w _j(\tau )\,, w _m(0)\rangle
\,=\,K_{jm}(|\tau |)\,\,\,,
\end{array}
\]
\[
K_{jm}(\tau)\,=\,\int_0^{\infty}\frac {e^{i\omega\tau}+
\exp(\hbar\omega/T)e^{-\,i\omega\tau}} {1+\exp(\hbar\omega/T)}\,
\,\sigma_{jm}(\omega)\,d\omega\,\,\,,
\]
where $\,\sigma_{jm}(\omega)\,$ is a non-negatively defined
real-valued spectrum matrix. This information about the random
sources $\,w(t)\,$ is sufficient in the framework of the Born
approximation or ``one-loop'' and ``ladder'' approximations and their
analogues (such as ``Bourret approximation'', ``Kraichnan
approximation'', etc. \cite{kl}). Moreover, if $\,w(t)\,$ possess
Gaussian statistics ($\,\mathcal{W}\,$ is ``Gaussian thermostat'')
then the spectrum $\,\sigma_{jm}(\omega)\,$ contains exhaustive
information about them, and principally one can exactly find any
average value. In all that cases, organization of the correlation
matrix $\,K_{jm}(\tau)\,$ automatically ensures \cite{i2,i5,i8}
balance of noise and dissipation (radiation and absorption of energy
by $\,\mathcal{W}\,$) in agreement with the fluctuation-dissipation
theorem. If $\,w(t)\,$ are non-Gaussian then they obey also
``generalized non-linear fluctuation-dissipation relations''
\cite{i8}.

{\bf 5}. Thus, the traditional approach to open systems based on
model kinetic equations (or equivalent ``Langevin equations'') has an
alternative in the form of exact linear Liouville and Schr\"{o}dinger
stochastic equations supplemented by exact universal expression for
characteristic functional of random sources (fluctuating parameters)
in these equations. At this alternative approach it is possible to
exploit concepts of wave functions and (pure) quantum states, their
superposition and unitary evolution. Correspondingly, mathematical
difficulties concentrate at statistical averaging of wave functions
and density matrices over the sources. This allows to use many
achievements of the theory of linear systems with randomly varying
parameters \cite{kl}. In this respect it is useful to notice that
non-Gaussian statistical models of fluctuating parameters sometimes
are much better suitable for exact calculations than Gaussian ones
\cite{kl,bk0}. Therefore development of simple Hamiltonian dynamical
models of non-Gaussian thermostats is of great interest. From the
other hand, as was shown by example applications of the new approach
\cite{i2,i4,i5,i6}, quite useful results can be obtained even without
full statistical information about the sources.

\end{document}